\documentclass[intlimits,twoside,a4paper]{article}

\usepackage[cp1251]{inputenc}
\usepackage[eqsecnum]{cmpj3}

\DeclareTextSymbolDefault{\textquotedbl}{T1}

\usepackage{bm}

\issue{2020}{23}{4}{43713}
\doinumber{10.5488/CMP.23.43713}

\title[Magnetoelastic properties of a spin-1/2 Ising-Heisenberg diamond chain]%
{Magnetoelastic properties of a spin-1/2 Ising-Heisenberg diamond chain in vicinity of a triple coexistence point}

\author[N. Ferreira \textsl{et al.}]{N. Ferreira\refaddr{label1},
        J. Torrico\refaddr{label2}, S.M. de Souza\refaddr{label1}, O. Rojas\refaddr{label1},  J. Stre\v{c}ka\refaddr{label3}}
\addresses{
\addr{label1} Departamento de F{\'\i}sica, Universidade Federal de Lavras CP 3037, 37200--900, Lavras --- MG, Brazil
\addr{label2}Departamento de F{\'\i}sica, Universidade Federal de Minas Gerais, C. P. 702, 30123--970, Belo Horizonte, Mg, Brazil
\addr{label3} Department of Theoretical Physics and Astrophysics, Faculty of Science, P. J. \v{S}af\'{a}rik University, Park Angelinum 9, 040 01 Ko\v{s}ice, Slovakia
}

\date{Received June 23, 2020, in final form September 9, 2020}

\begin{document}
\maketitle

\begin{abstract}
We study magnetoelastic properties of a spin-1/2 Ising-Heisenberg
diamond chain, whose elementary unit cell consists of two decorating
Heisenberg spins and one nodal Ising spin. It is assumed that each
couple of the decorating atoms including the Heisenberg spins harmonically
vibrates perpendicularly to the chain axis, while the nodal atoms
involving the Ising spins are placed at rigid positions when ignoring
their lattice vibrations. An effect of the magnetoelastic coupling
on a ground state and finite-temperature properties is particularly
investigated close to a triple coexistence point depending on a spring-stiffness
constant ascribed to the Heisenberg interaction. The magnetoelastic
nature of the Heisenberg dimers is reflected through a non-null plateau
of the entropy emergent in a low-temperature region, whereas the specific
heat displays an anomalous peak slightly below the temperature region
corresponding to the entropy plateau. The magnetization also exhibits
a plateau in the same temperature region at almost saturated value
before it gradually tends to zero upon increasing of temperature.
The magnetic susceptibility displays within the plateau region an
inverse temperature dependence, which slightly drops above this plateau,
whereas an inverse temperature dependence is repeatedly recovered
at high enough temperatures.

\end{abstract}

\section{Introduction}

Atomic vibrations of the crystalline materials may influence the magnetic
ordering and vice versa. This effect usually has peculiar manifestations
especially in a close vicinity of phase transitions related to a breakdown
of spontaneous long-range order of two- and three-dimensional magnetic
crystals \cite{Henriques,massimino,boubcheur}. The magnetoelastic
interaction produces deformation of a lattice structure (magnetostriction)
when applying the external magnetic field and, consequently, magneto-thermodynamic
properties of ferromagnetic materials are also altered. A rigorous
thermodynamic study of the magnetic crystals involving their magnetic,
vibrational, and elastic properties still remains a challenging problem
of current research interest owing to computational difficulties arising
from a mutual coupling of the magnetic and lattice degrees of freedom
through the magnetoelastic interaction. The magnetoelastic changes
are typically quite small, the measured strain is for instance of the
order of $10^{-5}-10^{-4}$ for Fe-, Ni- and Co-based alloys although
some specific materials like $\mathrm{Tb_{0.3}Dy_{0.7}Fe_{1.9}}$
may exhibit giant magnetoelastic changes with the measured strain
of the order $\sim10^{-3}$ \cite{Armstrong}. A new type of magnetostriction
was found in the materials later referred to as ferromagnetic shape-memory
alloys such as $\mathrm{Ni_{2}MnGa}$ \cite{tickle,tickle2}.

An effect of the magnetoelastic coupling was previously investigated
in a class of the mixed spin-(1/2,$S$) Ising models on decorated
planar lattices \cite{strecka12,strecka19}. Magnetic and lattice
degrees of freedom were in this particular case decoupled from each
other through the local canonical transformation \cite{enting}, which
either gives rise to an effective next-nearest-neighbor interaction
for the spin case $S=1/2$ \cite{strecka12} or an effective three-site
four-spin interaction and uniaxial single-ion anisotropy for the spin
case $S>1/2$~\cite{strecka19}. It was evidenced that the magnetoelastic
coupling enforces a remarkable spin frustration of the decorating
atoms, which was comprehensively studied in the mixed-spin Ising
model with the three-site four-spin interaction on decorated planar
lattices \cite{jascur16,stubna17} with the help of exact mapping
transformations \cite{fis59,syo72,roj09,str10,roj11}.

On the other hand, the magnetic behavior of one-dimensional Ising
systems drew attention due to the influence of the lattice compressibility.
In early 1960-ies Mattis and Schultz \cite{mattis} reported an exact
solution for the compressible Ising chain with free boundary condition
and concluded that there is no effect due to the spin-lattice coupling.
Later Enting \cite{enting} considered the periodic boundary condition
and verified that the effective spin Hamiltonian is equivalent to
a rigid Ising chain with first- and second-neighbor interactions.
Salinas \cite{salinas} obtained the free energy of the compressible
Ising chain subjected to fixed forces by a standard Legendre transformation,
which relates it to the free energy of the compressible Ising chain
confined to a fixed length. Early in 1980s  Kne\v{z}evi\v{c}
and Milo\v{s}evi\v{c} \cite{knezevic} considered the compressible
Ising chain with higher spin values $S=1$ and $S=3/2$, which can
be mapped to an effective rigid spin Hamiltonian with an additional
biquadratic interaction.

More recently, several aspects of compressible spin chains were investigated
such as the effect of two independent fields on the compressible
Ising chain \cite{Lemos}, thermodynamic properties of the Ising-chain
model accounting both for elastic and vibrational degrees of freedom
\cite{karol} and the magnetocaloric properties of the Ising chain
\cite{Yan}. The seminal contribution in this field of study  was by Derzhko and co-workers when rigorously solving a set of four
deformable spin-chain models \cite{Oleg}. Among other matters, Derzhko
et al. proved that the (inverse) compressibility of the Ising
chain in a longitudinal field and the quantum $XX$ chain in a transverse
field shows a sudden jump at field-driven quantum phase transition,
while it gradually diminishes near quantum critical points of the
Ising chain in a transverse field and the Heisenberg-Ising chain \cite{Oleg}.

Lately, different versions of the Ising-Heisenberg diamond chains
have provided a useful playground full of intriguing features and
unexpected findings such as the existence of intermediate
magnetization plateaus \cite{lis14,lis14_2,nerses-14}, Lyapunov exponent
and superstability \cite{nerses-13}, the non-conserved magnetization
and ``fire-and-ice'' ground states \cite{jor-18},
the enhanced magnetocaloric effect \cite{Galisova-14}, the pseudo-critical
behavior mimicking a temperature-driven phase transition \cite{sou17,gal15,tor16,Isaac}
or the pseudo-universality \cite{univers18}. Most importantly, Derzhko
and co-workers \cite{oleg-15} convincingly evidenced that the exact
solution for the Ising-Heisenberg diamond chain may be used as a useful
starting point for the perturbative treatment of the full Heisenberg
counterpart model. It was shown that this type of many-body perturbation
theory may even bring insight into exotic quantum states such as a
quantum spin liquid not captured by the original Ising-Heisenberg
model \cite{oleg-15}.

This article is organized as follows. Section~2 is devoted to a definition
and solution of the spin-1/2 Ising-Heisenberg diamond chain with the
vibrating character of the Heisenberg dimers. The ground-state phase
diagram as a function of the spring stiffness, magnetoelastic constant
and geometric structure is explored in section~3. In section~4 we present
the thermodynamics of the model, where the magnetization, entropy
and specific heat are analyzed in detail. Finally our conclusions
are reported in section~5.

\section{Ising-Heisenberg diamond chain with magnetoelastic coupling}

\begin{figure}[!t]
\centering \includegraphics[width=0.8\textwidth]{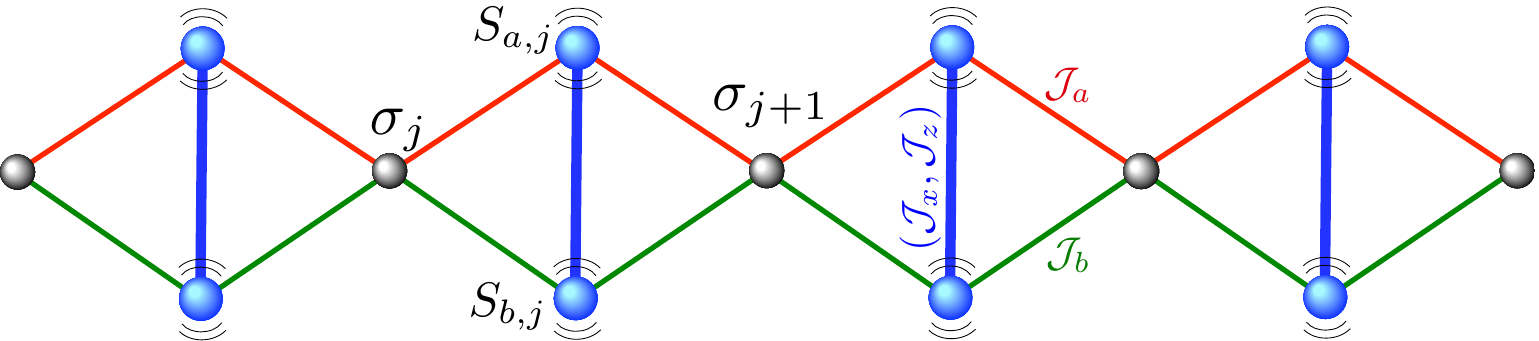}
\caption{(Colour online) A schematic representation of the spin-1/2 Ising-Heisenberg
diamond chain with the magnetoelastic coupling. The Ising spins $\sigma_{j}$
are placed at rigid lattice positions, while the Heisenberg spins
$S_{a,j}$ and $S_{b,j}$ harmonically vibrate in a direction perpendicular
with respect to the chain axis.}
\label{fig:Sdmd-chn}
\end{figure}

Let us consider the spin-1/2 Ising-Heisenberg diamond chain schematically
depicted in figure~\ref{fig:Sdmd-chn}, which  in an elementary
unit cell involves two Heisenberg spins $S_{a,j}$ and $S_{b,j}$ and one nodal
Ising spin $\sigma_{j}$. It will be further assumed that the decorating
atoms involving the Heisenberg spins harmonically vibrate perpendicularly
to the chain axis, while the nodal atoms involving the Ising spins
are rather rigid when neglecting their lattice vibrations.

Under this condition, the spin-1/2 Ising-Heisenberg diamond chain
can be defined through the Hamiltonian
\begin{equation}
{\cal{H}}=\sum_{j=1}^{N}{\cal H}_{j}=\sum_{j=1}^{N}\left({\cal{H}}_{j}^{({\text p})}+{\cal{H}}_{j}^{({\text{me}})}\right),
\label{eq:H-tot}
\end{equation}
where ${\cal H}_{j}^{({\rm p})}$ corresponds to the pure phonon part
and ${\cal H}_{j}^{({\rm me})}$ stands for the magnetoelastic part
of the Hamiltonian ${\cal H}_{j}$ explicitly given in what follows.

\begin{figure}[ht]
	\hspace{5mm}
	\includegraphics[width=0.4\textwidth]{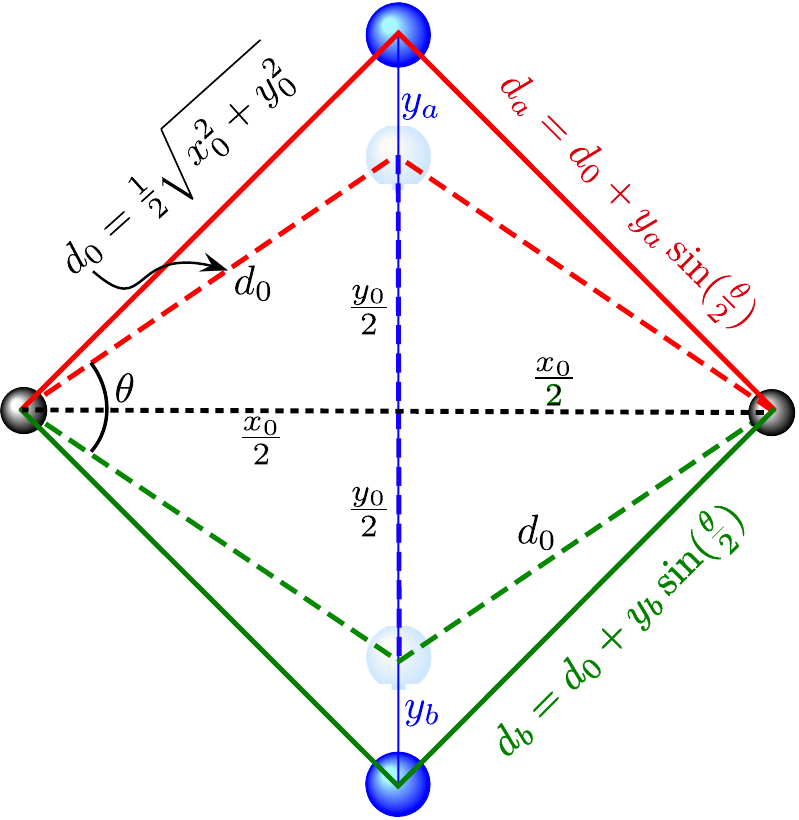}%
	\hfill%
	\includegraphics[width=0.4\textwidth]{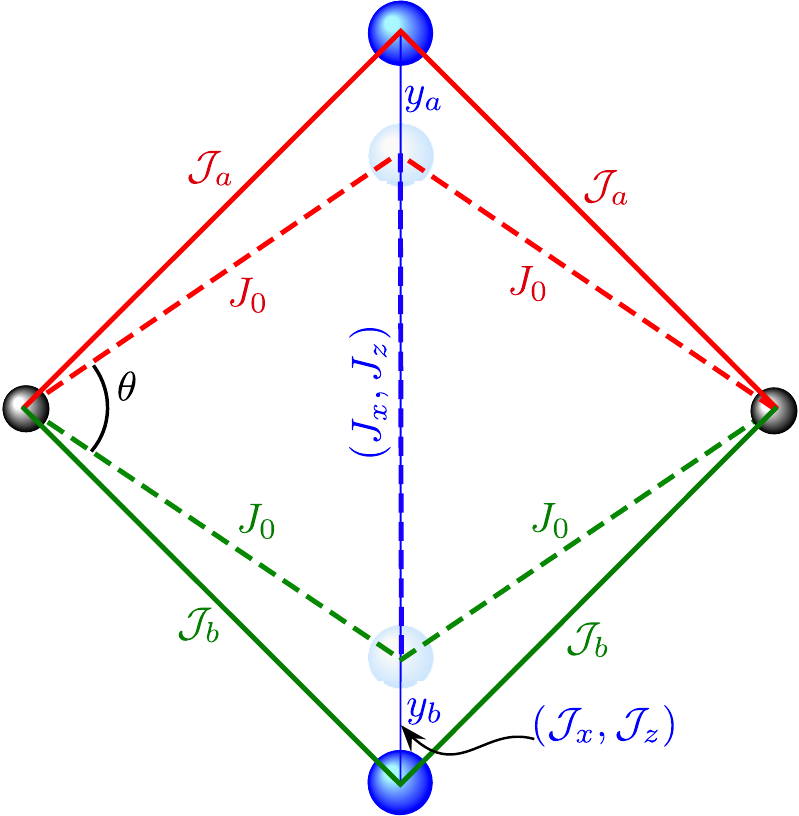}%
		\hspace{5mm}
	\\%
	\parbox[t]{0.48\textwidth}{%
		\centerline{(a)}%
	}%
	\hfill%
	\parbox[t]{0.48\textwidth}{%
		\centerline{(b)}%
	}%
\caption{(Colour online) A specification of the diamond unit cell under
	the geometric deformation through the displacements (a) and the exchange
	interactions (b).}
\label{fig:Dmd-dfrm}
\end{figure}

A specification of the displacements within the diamond unit cell
is depicted in figure~\ref{fig:Dmd-dfrm}~(a), where $x_{0}$ is the
equilibrium distance between the Ising spins, $y_{0}$ corresponds
to the equilibrium distance between the Heisenberg spins, and $d_{0}$
is the equilibrium distance between the Ising and Heisenberg spins.
It is supposed that the decorating atoms $a$ and $b$ including the
Heisenberg spins  perform harmonic oscillations around their
equilibrium lattice positions, which can be characterized through
small displacements $y_{a}$ and $y_{b}$, while nodal Ising spins
are considered at a rigid position (heavy atoms), this assumption
is also reasonable because there is no direct interaction between
Ising spins. Consequently, the instantaneous distances between the
Heisenberg and Ising spins are changed to $d_{a}=d_{0}+y_{a}\sin({\theta}/{2})$
and $d_{b}=d_{0}+y_{b}\sin({\theta}/{2})$ though the distance
$x_{0}$ between the Ising spins remains unaltered. Note that $y_{a}$
and $y_{b}$ are considered positive when dimer particles  expand,
but when they compress, we consider them negative.

Under the linear approximation, the magnetoelastic part of the bond
Hamiltonian (\ref{eq:H-tot}) can be written in this form
\begin{alignat}{1}
{\cal H}_{j}^{({\rm me})}= & -\mathcal{J}_{x,j}\left(S_{a,j}^{x}S_{b,j}^{x}+S_{a,j}^{y}S_{b,j}^{y}\right)-\mathcal{J}_{z,j}S_{a,j}^{z}S_{b,j}^{z}-\left(\mathcal{J}_{a,j}S_{a,j}^{z}+\mathcal{J}_{b,j}S_{b,j}^{z}\right)\left(\sigma_{j}+\sigma_{j+1}\right)\nonumber \\
 & -h_{\mathrm{H}}\left(S_{a,j}^{z}+S_{b,j}^{z}\right)-\frac{h_{\mathrm{I}}}{2}\left(\sigma_{j}+\sigma_{j+1}\right),\label{eq:H-me}
\end{alignat}
where the exchange interactions {[}see figure~\ref{fig:Dmd-dfrm}~(b){]}
are given by
\begin{eqnarray}
\mathcal{J}_{x,j}\!\!\! & = & \!\!\!J_{x}\left[1-\kappa(\mathsf{y}_{a,j}+\mathsf{y}_{b,j})\right],\qquad\mathcal{J}_{z,j}=J_{z}\left[1-\kappa(\mathsf{y}_{a,j}+\mathsf{y}_{b,j})\right],\nonumber \\
\mathcal{J}_{a,j}\!\!\! & = & \!\!\!J_{0}\left[1-\eta\mathsf{y}_{a,j}\sin\left(\frac{\theta}{2}\right)\right],\qquad\mathcal{J}_{b,j}=J_{0}\left[1-\eta\mathsf{y}_{b,j}\sin\left(\frac{\theta}{2}\right)\right].\label{eq:Jb}
\end{eqnarray}
Here, $J_{x}$ and $J_{z}$ correspond to $xy$- and $z$-component
of the Heisenberg exchange interaction when assuming the decorating
atoms at equilibrium positions. Similarly, $J_{0}$ corresponds to
the Ising exchange interaction between the Heisenberg and Ising spins
at equilibrium positions. Finally, $\kappa$ is the linear expansion
coefficient for the magnetoelastic coupling within the Heisenberg
dimers and $\eta$ is the linear expansion coefficient for the magnetoelastic
coupling between the Ising and Heisenberg spins. We do not include
a term of second-order contribution, because we  assume simply a
linear dependence as considered in reference \cite{strecka19} and in 
references therein.

For simplicity, from now on, we will consider the standard atomic
units (au) which means the Planck's constant is $\hbar=1$, Boltzmann
constant becomes $k_{\mathrm{B}}=1$, Bohr magneton constant is $\mu_{\mathrm{B}}=1/2$
and the gyromagnetic ratio is estimated as $\gamma\approx2$, so we
have $\mu_{\mathrm{B}}\gamma=1$. Therefore, exchange interaction parameters,
external magnetic field, displacements, Hooke's constant are all in
atomic units.

The purely elastic part of the bond Hamiltonian (\ref{eq:H-tot})
can be defined as follows:
\begin{alignat}{1}
{\cal H}_{j}^{({\rm p})}= & \frac{\mathsf{p}_{a,j}^{2}}{2m}\!+\!\frac{\mathsf{p}_{b,j}^{2}}{2m}\!+\!\frac{\bar{K}}{2}\negmedspace\left(\mathsf{y}_{a,j}^{2}\!+\!\mathsf{y}_{b,j}^{2}\right)\!+\!\frac{\mathsf{k}_{_{\mathrm{H}}}}{2}\left(\mathsf{y}_{a,j}\!+\!\mathsf{y}_{b,j}\right)^{2},\label{eq:H-ph}
\end{alignat}
where $\mathsf{p}_{a,j}$ and $\mathsf{p}_{b,j}$ are momenta of the
decorating atoms with the mass $m$, $\mathsf{y}_{a,j}$ and $\mathsf{y}_{b,j}$
denote their displacements from equilibrium positions, $\mathsf{k}_{_{\mathrm{H}}}$
is the ``spring-stiffness'' constant ascribed to the Heisenberg
coupling, and $\bar{K}=2\mathsf{k}{}_{_{\mathrm{I}}}\sin^{2}({\theta}/{2})$
is the effective ``spring-stiffness'' constant of the Ising coupling
when $\mathsf{k}{}_{_{\mathrm{I}}}$ is a true ``spring-stiffness'' constant
ascribed to the Ising coupling.

\subsection{Local canonical transformation}

The Hamiltonian (\ref{eq:H-tot}) involves magnetoelastic and pure
elastic (phonon) contributions, which are coupled together through
the linear expansion coefficients $\kappa$ and $\eta$ pertinent
to the magnetoelastic couplings. However, both contributions can be
decoupled through the local canonical coordinate transformation
\begin{equation}
\boldsymbol{q}_{j}=\frac{1}{\sqrt{2}}\left(\mathsf{y}_{a,j}+\mathsf{y}_{b,j}\right)\quad\text{and}\quad\bar{\boldsymbol{q}}_{j}=\frac{1}{\sqrt{2}}\left(\mathsf{y}_{a,j}-\mathsf{y}_{b,j}\right),\label{cct}
\end{equation}
which defines the positions of two fictitious particles. Analogously,
the momenta in the canonical coordinates take the form
\begin{equation}
\boldsymbol{p}_{j}=\frac{1}{\sqrt{2}}\left(\mathsf{p}_{a,j}+\mathsf{p}_{b,j}\right)\quad\text{and}\quad\bar{\boldsymbol{p}}_{j}=\frac{1}{\sqrt{2}}\left(\mathsf{p}_{a,j}-\mathsf{p}_{b,j}\right).\label{cmt}
\end{equation}
 Thus, the Hamiltonian (\ref{eq:H-ph}) in the canonical coordinates
can be rewritten as follows:
\begin{alignat}{1}
{\cal H}_{j}^{({\rm p})}= & \frac{\boldsymbol{p}_{j}^{2}}{2m}+\frac{\bar{\boldsymbol{p}}_{j}^{2}}{2m}+\frac{\bar{K}}{2}\left(\boldsymbol{q}_{j}^{2}+\bar{\boldsymbol{q}}_{j}^{2}\right)+\mathsf{k}_{_{\mathrm{H}}}\boldsymbol{q}_{j}^{2}\,.\label{eq:H-ph-c}
\end{alignat}

\subsection{Diagonalization of the magnetoelastic part}

Since the commutation relation $[{\cal H}_{i}^{({\rm me})},{\cal H}_{j}^{({\rm me})}]=0$
is satisfied, the magnetoelastic part of the bond Hamiltonian (\ref{eq:H-me})
can be diagonalized separately for each unit cell and the respective
eigenvalues can be expressed as a function of the canonical coordinate
for a position $\boldsymbol{q}_{j}$ in the following form
\begin{alignat}{1}
\varepsilon_{k,j}= & \mathfrak{e}_{k,j}^{(0)}+\mathfrak{e}_{k,j}^{(1)}\boldsymbol{q}_{j}\,,\label{eq:ek}
\end{alignat}
with the first index being $k=\{1,2,3,4\}$ and the second index specifying the
unit cell. The eigenvalues~(\ref{eq:ek}) were decomposed into two
terms, whereas the first terms $\mathfrak{e}_{k,j}^{(0)}$ correspond
to a fully rigid diamond chain
\begin{eqnarray}
\mathfrak{e}_{1,j}^{(0)}\!\!\! & = & \!\!\!-\frac{J_{z}}{4}-h_{\mathrm{H}}-\left(\frac{h_{\mathrm{I}}}{2}+J_{0}\right)\mu_{j}\,,\qquad\mathfrak{e}_{2,j}^{(0)}=-\frac{J_{z}}{4}+h_{\mathrm{H}}-\left(\frac{h_{\mathrm{I}}}{2}-J_{0}\right)\mu_{j}\,,\nonumber \\
\mathfrak{e}_{3,j}^{(0)}\!\!\! & = & \!\!\!\frac{J_{z}}{4}+\frac{J_{x}}{2}-\frac{h_{\mathrm{I}}}{2}\mu_{j}\,,\qquad\qquad\qquad\mathfrak{e}_{4,j}^{(0)}=\frac{J_{z}}{4}-\frac{J_{x}}{2}-\frac{h_{\mathrm{I}}}{2}\mu_{j}\,,\label{eq:e04}
\end{eqnarray}
which can be alternatively interpreted as the energy eigenvalues when
the decorating atoms are at equilibrium positions $y_{a,j}=y_{b,j}=0$
and for simplicity, we  denoted $\mu_{j}=(\sigma_{j}+\sigma_{j+1})$.
The second terms $\mathfrak{e}_{k,j}^{(1)}$ determine a vibrational
contribution to the overall energy, which is given by
\begin{eqnarray}
\mathfrak{e}_{1,j}^{(1)}\!\!\! & = & \!\!\!\frac{\sqrt{2}}{4}\left[{\it J_{z}}\kappa+2J_{0}\eta\mu_{j}\sin\left(\frac{\theta}{2}\right)\right],\qquad\mathfrak{e}_{2,j}^{(1)}=\frac{\sqrt{2}}{4}\left[{\it J_{z}}\kappa-2J_{0}\eta\mu_{j}\sin\left(\frac{\theta}{2}\right)\right],\nonumber \\
\mathfrak{e}_{3,j}^{(1)}\!\!\! & = & \!\!\!-\frac{\sqrt{2}}{4}\kappa(2J_{x}+J_{z}),\qquad\qquad\qquad\mathfrak{e}_{4,j}^{(1)}=\frac{\sqrt{2}}{4}\kappa(2J_{x}-J_{z}).\label{eq:e14}
\end{eqnarray}
The eigenvectors, which correspond to the energy eigenvalues (\ref{eq:ek}),
can be expressed using the notation $|_{S_{b}^{z}}^{S_{a}^{z}}\rangle_{j}$
as follows
\begin{eqnarray*}
|\varphi_{1}\rangle_{j}=|_{+}^{+}\rangle_{j}\,,\quad|\varphi_{2}\rangle_{j}&=&|_{-}^{-}\rangle_{j}\,,\quad|\varphi_{3}\rangle_{j}=\sin(\vartheta_{j})|{}_{-}^{+}\rangle_{j}-\cos(\vartheta_{j})|{}_{+}^{-}\rangle_{j}\,,\\[2ex]
|\varphi_{4}\rangle_{j}&=&\cos(\vartheta_{j})|{}_{-}^{+}\rangle_{j}+\sin(\vartheta_{j})|{}_{+}^{-}\rangle_{j}\,,
\end{eqnarray*}
whereas the relevant mixing angle $\vartheta_{j}$ entering  the
last two eigenvectors is defined as follows:
\[
\tan(2\vartheta_{j})=\frac{J_{x}(1-\kappa\sqrt{2}q_{j})}{\sqrt{2}J_{0}\eta\sin(\frac{\theta}{2})\bar{q}_{j}\mu_{j}}\,.
\]
It is worth to mention that the last two eigenvectors depend on the
positions $q_{j}$ and $\bar{q}_{j}$ of the decorating atoms.

\subsection{Diagonalization of the phonon part}

After performing the canonical coordinate transformation and diagonalizing,
the magnetoelastic part for each eigenvalue of bond
Hamiltonian takes the form
\begin{equation}
{\cal H}_{k,j}=\mathfrak{e}_{k,j}^{(0)}+\mathfrak{e}_{k,j}^{(1)}\boldsymbol{q}_{j}+\frac{\boldsymbol{p}_{j}^{2}}{2m}+\frac{\bar{\boldsymbol{p}}_{j}^{2}}{2m}+\frac{\bar{K}}{2}\left(\boldsymbol{q}_{j}^{2}+\bar{\boldsymbol{q}}_{j}^{2}\right)+\mathsf{k}_{_{\mathrm{H}}}\boldsymbol{q}_{j}^{2}\,.\label{eq:H-cpl-Hmt}
\end{equation}
The  above result can be subsequently fully decoupled
and diagonalized by completing square through an additional local
transformation for relative position of the Heisenberg dimers
\begin{equation}
\boldsymbol{q}_{j}=\boldsymbol{q}_{j}^{\prime}-\frac{\mathfrak{e}_{k,j}^{(1)}}{K}\,,\label{eq:q'-q}
\end{equation}
which is defined through the effective spring-stiffness constants
$K=\bar{K}+2\mathsf{k}_{_{\mathrm{H}}}$ and $\bar{K}=2\mathsf{k}_{_{\mathrm{I}}}\sin^{2}({\theta}/{2})$.
Substituting a shift of the canonical coordinate for position (\ref{eq:q'-q})
into equation~(\ref{eq:H-cpl-Hmt}), one actually achieves a decoupling
of the magnetoelastic and pure phonon parts of the bond Hamiltonian,
whereas the effective phonon part becomes
\begin{alignat}{1}
\mathcal{H}_{j}^{({\rm p})}= & \frac{\boldsymbol{p}_{j}^{2}}{2m}+\frac{\bar{\boldsymbol{p}}_{j}^{2}}{2m}+\frac{K}{2}(\boldsymbol{q}_{j}^{\prime})^{2}+\frac{\bar{K}}{2}\bar{\boldsymbol{q}}_{j}^{2}\,,\label{eq:H-ph-r}
\end{alignat}
while the effective magnetoelastic part reads
\begin{alignat}{1}
\mathcal{H}_{k,j}^{({\rm me})}= & \mathfrak{e}_{k,j}^{(0)}-\frac{\bigl(\mathfrak{e}_{k,j}^{(1)}\bigr)^{2}}{2K}\,.\label{eq:H-ph-n-1}
\end{alignat}

To proceed further, let us introduce the annihilation $\boldsymbol{b}_{j}$
and creation $\boldsymbol{b}_{j}^{\dagger}$ bosonic operators describing
phonon positions, which satisfy the standard bosonic commutation relations
$[\boldsymbol{b}_{j},\boldsymbol{b}_{j'}^{\dagger}]=\delta_{j,j'}$
and $[\boldsymbol{b}_{j},\boldsymbol{b}_{j'}]=0$. The shifted position
$\boldsymbol{q}'_{j}$ and momentum operator $\boldsymbol{p}_{j}$
of the Heisenberg dimer can be consequently rewritten in terms of
the newly defined creation $\boldsymbol{b}_{j}^{\dagger}$ and annihilation
$\boldsymbol{b}_{j}$ bosonic operators (in units of $\hbar=1$)
\begin{equation}
\boldsymbol{q}'_{j}=\frac{1}{\sqrt{2m\omega}}(\boldsymbol{b}_{j}^{\dagger}+\boldsymbol{b}_{j})\quad\text{and}\quad\boldsymbol{p}_{j}=i\sqrt{\frac{m\omega}{2}}(\boldsymbol{b}_{j}^{\dagger}-\boldsymbol{b}_{j}),
\end{equation}
where $\omega=\sqrt{K/m}$ is the respective phonon frequency. Similarly,
one may also introduce the annihilation $\bar{\boldsymbol{b}}_{j}$
and creation $\bar{\boldsymbol{b}}_{j}^{\dagger}$ bosonic operators
describing the corresponding phonon term, which also satisfy the standard
bosonic commutation relations $[\bar{\boldsymbol{b}}_{j},\bar{\boldsymbol{b}}_{j'}^{\dagger}]=\delta_{j,j'}$
and $[\bar{\boldsymbol{b}}_{j},\bar{\boldsymbol{b}}_{j'}]=0$. Therefore,
the position $\bar{\boldsymbol{q}}_{j}$ and momentum $\bar{\boldsymbol{p}}_{j}$
can be also defined as
\begin{equation}
\bar{\boldsymbol{q}}_{j}=\frac{1}{\sqrt{2m\bar{\omega}}}(\bar{\boldsymbol{b}}_{j}^{\dagger}+\bar{\boldsymbol{b}}_{j})\quad\text{and}\quad\bar{\boldsymbol{p}}_{j}=i\sqrt{\frac{m\bar{\omega}}{2}}(\bar{\boldsymbol{b}}_{j}^{\dagger}-\bar{\boldsymbol{b}}_{j}),
\end{equation}
where $\bar{\omega}=\sqrt{\bar{K}/m}$ is the respective phonon frequency.
The phonon part of the Hamiltonian (\ref{eq:H-ph-r}) can be subsequently
expressed in terms of the number operators $\boldsymbol{n}_{j}$ and
$\bar{\boldsymbol{n}}_{j}$ for two aforedescribed phonons
\begin{alignat}{1}
\mathcal{H}_{j}^{({\rm p})}= & \left(\frac{1}{2}\!+\!\boldsymbol{b}_{j}^{\dagger}\boldsymbol{b}_{j}\right)\omega\!+\!\left(\frac{1}{2}\!+\!\bar{\boldsymbol{b}}_{j}^{\dagger}\bar{\boldsymbol{b}}_{j}\right)\bar{\omega}=\left(\frac{1}{2}\!+\!\boldsymbol{n}_{j}\right)\omega\!+\!\left(\frac{1}{2}\!+\!\bar{\boldsymbol{n}}_{j}\right)\bar{\omega}.\label{eq:H-ph-n}
\end{alignat}
In this way, we have achieved diagonalization of the phonon part of
the Hamiltonian (\ref{eq:H-ph-n}) as the number operators $\boldsymbol{n}_{j}$
and $\bar{\boldsymbol{n}}_{j}$ acquire the following set of eigenvalues
$n_{j}$ and $\bar{n}_{j}\in\{0,1,2,...\}$ with regard to the bosonic
character of the underlying operators. It is worthwhile to remark
that the bond Hamiltonian $\mathcal{H}_{k,j}=\mathcal{H}_{k,j}^{(\mathrm{me})}+\mathcal{H}_{j}^{(\mathrm{p})}$
is now expressed in a fully diagonal form with regard to the diagonal
character of the magnetoelastic (\ref{eq:H-ph-n-1}) and the phonon
(\ref{eq:H-ph-n}) parts, which additionally mutually commute with
each other, which will be of principal importance for calculation of
the partition function presented in section~4.

\section{Ground-state phase diagram}

Before exploring the magnetoelastic properties, let us  first analyze
a ground-state phase diagram of the spin-1/2 Ising-Heisenberg diamond
chain supplemented with the magnetoelastic coupling, which exhibits
three different ground states on the assumption that $k_{\mathrm{H}}=k_{\mathrm{I}}$ and
$\kappa=\eta$. The first ground state can be classified as the saturated
paramagnetic state (SA) given by the eigenvector
\begin{equation}
|\mathrm{SA}\rangle=\prod_{j=1}^{N}|_{+}^{+}\rangle_{j}|\uparrow\rangle_{j}\,,
\end{equation}
whereas the former (latter) state vector defines a spin state of the
Heisenberg dimer (the Ising spin) from the $j$th unit cell. The saturated
paramagnetic state has the following eigenenergy
\begin{equation}
E_{\mathrm{SA}}=-\frac{J_{z}}{4}-h_{\mathrm{H}}-\frac{h_{\mathrm{I}}}{2}-J_{0}-\frac{\left({\it J_{z}}\kappa+2J_{0}\eta\sin\frac{\theta}{2}\right)^{2}}{16K}\,.
\end{equation}
The second ground state can be viewed as the classical ferrimagnetic
phase (FI$_{1}$) given by the eigenvector
\begin{equation}
|\mathrm{FI}_{1}\rangle=\prod_{j=1}^{N}|_{+}^{+}\rangle_{j}|\downarrow\rangle_{j}\,,
\end{equation}
whereas the corresponding energy becomes
\begin{equation}
E_{\mathrm{FI}_{1}}=-\frac{J_{z}}{4}-h_{\mathrm{H}}+\frac{h_{\mathrm{I}}}{2}+J_{0}-\frac{\left({\it J_{z}}\kappa-2J_{0}\eta\sin\frac{\theta}{2}\right)^{2}}{16K}\,.
\end{equation}
The sublattice magnetization of the Ising spins is $M_{\mathrm{I}}=-1/2$ per
unit cell, the sublattice magnetization of the Heisenberg spins is
$M_{\mathrm{H}}=1$ per unit cell so that the total magnetization per unit
cell is $M_{\mathrm{t}}=1/2$. The third ground state can be classified as
the quantum ferrimagnetic phase (FI$_{2}$) given by the eigenvector
\begin{equation}
|\mathrm{FI}_{2}\rangle=\prod_{j=1}^{N}\left(\cos(\vartheta_{j})|{}_{-}^{+}\rangle_{j}+\sin(\vartheta_{j})|{}_{+}^{-}\rangle_{j}\right)|\uparrow\rangle_{j}\,,
\end{equation}
whose corresponding eigenenergy reads as follows:
\begin{equation}
E_{\mathrm{FI}_{2}}=\frac{J_{z}}{4}-\frac{J_{x}}{2}-\frac{h_{\mathrm{I}}}{2}-\frac{\kappa^{2}(2J_{x}-J_{z})^{2}}{16K}\,.
\end{equation}
The sublattice magnetization of the Heisenberg spins is $M_{\mathrm{H}}=0$
due to a singlet-like character of the Heisenberg dimers and hence,
the sublattice magnetization of the Ising spins $M_{\mathrm{I}}=1/2$ provides
the only nonzero contribution to the total magnetization per unit
cell $M_{\mathrm{t}}=1/2$.

\begin{figure}[h]
\centering \includegraphics[width=0.98\textwidth]{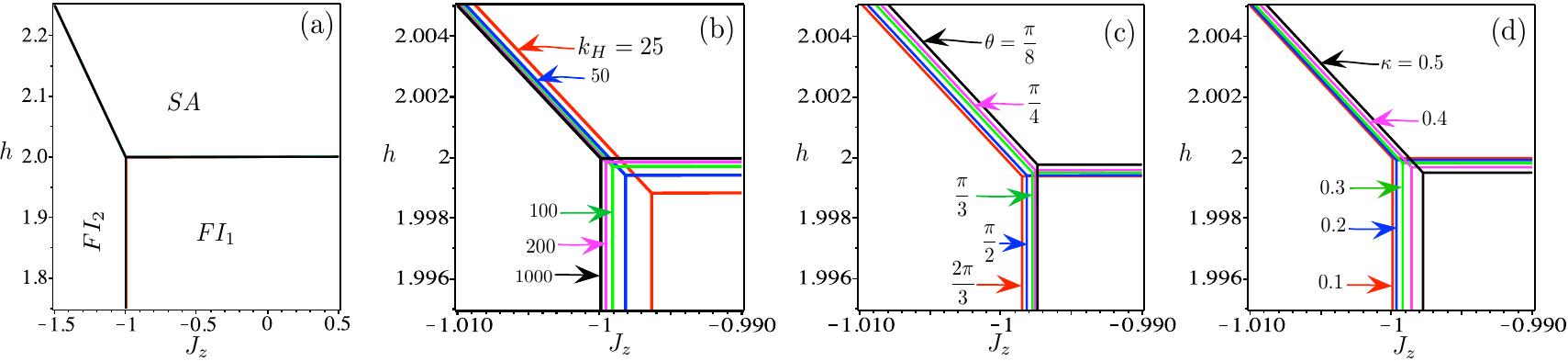}
\caption{(Colour online) \label{fig:Phase-diagram} Ground-state phase diagram in $J_{z}-h$
plane for $J_{0}=-1$, $J_{x}=1$ and: (a)--(b) $\kappa=0.5$, $\theta=\piup/2$
and $k_{\mathrm{H}}=\{25,50,200,1000\}$; (c) $\kappa=0.5$, $k_{\mathrm{H}}=50$ and
$\theta=\{2\piup/3,\piup/2,\piup/3,\piup/4,\piup/8\}$; (d) $\theta=\piup/3$,
$k_{\mathrm{H}}=50$ and $\kappa=\{0.1,0.2,0.3,0.4,0.5\}$.}
\end{figure}

Here, we consider the local magnetic fields $h=h_{\mathrm{I}}=h_{\mathrm{H}}$ acting
on the Ising and Heisenberg spins, which physically corresponds to
the equality of their Land\'{e} g-factors. A few typical ground-state
phase diagrams are plotted in figure~\ref{fig:Phase-diagram} in $J_{z}-h$
plane for the fixed values of the coupling constants $J_{0}=-1$ and
$J_{x}=1$. It is evident that the ground-state phase boundaries are
only gradually shifted with respect to a perfectly rigid limit when
assuming realistic (rather high) values of the spring-stiffness
constants (see \cite{can06} for a comparison). As a matter of
fact, the changes in the ground-state phase diagram shown in figure~\ref{fig:Phase-diagram}~(a)
for the fixed values of $\kappa=0.5$ and $\theta=\piup/2$ due to variation
of the spring-stiffness constant $k_{\mathrm{H}}=\{25,50,200,1000\}$ are almost
indistinguishable within the displayed scale, while they become evident
just in a magnified scale as illustrated in figure~\ref{fig:Phase-diagram}~(b).
Note that  similar effects also result from the changes of other interaction
parameters {[}see figure~\ref{fig:Phase-diagram}~(c)--(d){]}. The role
of lattice geometry can be traced back in figure~\ref{fig:Phase-diagram}~(c),
where the ground-state phase diagrams are plotted for fixed values
of $\kappa=0.5$, $k_{\mathrm{H}}=50$ and several values of the angle $\theta=\{2\piup/3,\piup/2,\piup/3,\piup/4,\piup/8\}$.
Finally, the effect of magnetoelastic constant on the ground-state
phase diagrams is illustrated in figure~\ref{fig:Phase-diagram}~(d)
when assuming a fixed value of $\theta=\piup/3$, $k_{\mathrm{H}}=50$ upon variation
of $\kappa=\{0.1,0.2,0.3,0.4,0.5\}$.

The phase boundary between two ferrimagnetic phases FI$_{1}$ and
FI$_{2}$ is given by
\begin{equation}
J_{z}=\frac{\left[4\!\left(2J_{0}\!+\!J_{x}\right)\!k_{\mathrm{H}}\!-\!J_{0}^{2}\kappa^{2}\right]\!\left(\sin^{2}\frac{\theta}{2}+1\right)\!+\!\left(J_{0}^{2}\!+\!J_{x}^{2}\right)\!\kappa^{2}}{\kappa^{2}\left(J_{x}-J_{0}\sin\frac{\theta}{2}\right)+4k_{\mathrm{H}}\left(\sin^{2}\frac{\theta}{2}+1\right)}\,,
\end{equation}
which is independent of the magnetic field $h$ as evidenced by a
vertical character of the relevant phase boundaries in figure~\ref{fig:Phase-diagram}.
The phase boundary between the phases FI$_{1}$ and SA follows
from the formula
\begin{equation}
h=-2J_{0}-\frac{J_{0}J_{z}\kappa^{2}\sin\frac{\theta}{2}}{4{\it k}_{\mathrm{H}}\left(\sin^{2}\frac{\theta}{2}+1\right)}\,.
\end{equation}
While this phase boundary is for a perfectly rigid model ($\kappa=0$)
completely independent of $J_{z}$, the model with nonzero magnetoelastic
coupling constant $\kappa\neq0$ shows a relatively weak dependence
on the coupling constant $J_{z}$ because $\kappa$ is in general a
very small quantity ($\kappa\ll1$), while the spring-stiffness constant
$k_{\mathrm{H}}\gg\mathfrak{e}_{k,j}^{(1)}$ {[}i.e. $k_{\mathrm{H}}\gg J_{0}J_{z}$,
see figure \ref{fig:Phase-diagram}~(b)--(d) for illustration{]}. A similar
finding concerns with the phase boundary between FI$_{2}$ and SA
phases, which is given by
\begin{alignat}{1}
h= & \frac{1}{2}\left(J_{x}-2J_{0}-J_{z}\right)+\kappa^{2}\frac{\left({\it J_{x}}+J_{0}\sin\frac{\theta}{2}\right)\left({\it J_{x}}-J_{z}-J_{0}\sin\frac{\theta}{2}\right)}{8k_{\mathrm{H}}\left(\sin^{2}\frac{\theta}{2}+1\right)}\,.
\end{alignat}
This phase boundary depends on the coupling constant $J_{z}$ even
in the perfectly rigid limit ($\kappa=0$), but there appears a small
correction to this dependence once nonzero magnetoelastic coupling
constant ($\kappa\neq0$) is taken into consideration.

\section{Thermodynamics}

As previously commented, the phonon $\mathcal{H}_{j}^{(\mathrm{p})}$
and magnetoelastic $\mathcal{H}_{k,j}^{(\mathrm{me})}$ parts of the
bond Hamiltonian [\ref{eq:H-ph-n-1} and \eqref{eq:H-ph-n}] commute
with each other. For this reason, the bond Hamiltonians corresponding
to two different unit cells also satisfy the commutation relation
$[{\cal H}_{j},{\cal H}_{j'}]=0$. The partition function of the spin-1/2
Ising-Heisenberg diamond chain accounting for the magnetoelastic interaction
can be accordingly obtained by using the transfer-matrix approach.
A decoupled character of the phonon and magnetoelastic parts of the
Hamiltonian allows one to factorize the partition function into a
product of two terms
\begin{equation}
\mathcal{Z}_{N}={\cal Z}_{N}^{(\mathrm{p})}{\cal Z}_{N}^{(\mathrm{me})},\label{pf}
\end{equation}
whereas the former one ${\cal Z}_{N}^{(\mathrm{p})}$ denotes the
phonon contribution and the latter one ${\cal Z}_{N}^{(\mathrm{me})}$
corresponds to the magnetoelastic contribution. It is noteworthy that
the phonons corresponding to $(\boldsymbol{p},\boldsymbol{q})$ and
$(\bar{\boldsymbol{p}},\bar{\boldsymbol{q}})$ of the Heisenberg dimers
are independent of each other and hence, the phonon part of the partition
function can be expressed more explicitly as follows
\begin{equation}
{\cal Z}_{N}^{(\mathrm{p})}=\left(u\bar{u}\right)^{N},\label{pfp}
\end{equation}
where individual contributions stemming from two kinds of phonons
involved in the Hamiltonian (\ref{eq:H-ph-n-1}) follow from a summation
over all accessible values of the quantum numbers $n_{j}$ and $\bar{n}_{j}$
\begin{eqnarray}
u=\sum_{n_{j}=0}^{\infty}{\rm e}^{-\beta\left(\frac{1}{2}+n_{j}\right)\omega}=\frac{1}{2{\rm sinh}\left(\frac{\beta\omega}{2}\right)},\qquad\bar{u}=\sum_{\bar{n}_{j}=0}^{\infty}{\rm e}^{-\beta\left(\frac{1}{2}+\bar{n}_{j}\right)\bar{\omega}}=\frac{1}{2{\rm sinh}\left(\frac{\beta\bar{\omega}}{2}\right)}\,.
\end{eqnarray}
The magnetoelastic part of the partition function can be calculated
using the transfer matrix
\begin{equation}
\boldsymbol{W}=\left(\begin{array}{cc}
w_{1} & w_{0}\\
w_{0} & w_{-1}
\end{array}\right),\label{tm}
\end{equation}
which involves the Boltzmann factors of the $j$th Heisenberg dimer
defined as follows:
\begin{equation}
w_{\mu_{j}}=\sum_{k=1}^{4}{\rm e}^{-\beta\bigl[\mathfrak{e}_{k,j}^{(0)}-\frac{1}{2K}\bigl(\mathfrak{e}_{k,j}^{(1)}\bigr)^{2}\bigr]}\,.\label{eq:W_Bltz}
\end{equation}
In the above, the energy contributions $\mathfrak{e}_{k,j}^{(0)}$ and
$\mathfrak{e}_{k,j}^{(1)}$ are given by equations~(\ref{eq:e04}) and
(\ref{eq:e14}), respectively. After a straightforward diagonalization
of the transfer matrix (\ref{tm}), one gets two eigenvalues
\begin{equation}
\lambda_{\pm}=\frac{1}{2}\left(w_{1}+w_{-1}\pm\sqrt{(w_{1}-w_{-1})^{2}+4w_{0}^{2}}\right).
\end{equation}
The magnetoelastic part of the partition function can be expressed
via the transfer-matrix eigenvalues
\begin{equation}
{\cal Z}_{N}^{(\mathrm{me})}=\lambda_{+}^{N}+\lambda_{-}^{N}\,.\label{pfme}
\end{equation}
At this stage, one may substitute the phonon and magnetoelastic parts
of the partition functions (\ref{pfp} and \eqref{pfme}) into equation~(\ref{pf}) in order to get the final result for the partition function
and the associated free energy, which in the thermodynamic limit reduces
to the form
\begin{equation}
f=-\frac{1}{\beta}\lim_{N\to\infty}\frac{1}{N}\ln{\cal Z}_{N}=-\frac{1}{\beta}\ln\left(u\bar{u}\right)-\frac{1}{\beta}\ln\left(\lambda_{+}\right).
\end{equation}
With the free energy in hand, we are able to discuss magnetoelastic
properties of the spin-1/2 Ising-Heisenberg diamond chain at nonzero
temperatures.

In what follows our particular attention will be devoted to the magnetoelastic
behavior at and near a triple point, where all three phases SA, FI$_{1}$
and FI$_{2}$ coexist together at zero temperature. In the case of
a completely rigid model ($\kappa=\eta=0$), the three phases coexist
together for fixed parameters $J_{0}=-1$, $J_{x}=1$, $J_{z}=-1$
when the magnetic field $h\rightarrow2$ {[}see in figure~\ref{fig:Phase-diagram}~(a){]}.
After some algebraic manipulations, one obtains the following asymptotic
expression for the free energy of the rigid model $f_{0}=-T\ln\left(\lambda_{+}\right)$
in the zero-temperature limit $T\rightarrow0$
\begin{alignat}{1}
f_{0}= & E_{0}-T\ln\left(\frac{3+\sqrt{5}}{2}\right)-\frac{\sqrt{5}}{10}\left(h_{\mathrm{I}}-2\right)-\frac{1}{2}\left(1+\frac{\sqrt{5}}{5}\right)\left(h_{\mathrm{H}}-2\right),\label{eq:f0T}
\end{alignat}
where $E_{0}=E_{\mathrm{FI}_{2}}=E_{\mathrm{FI}_{1}}=E_{\mathrm{SA}}$ defines the corresponding
ground-state energy at a triple point. Now, one may differentiate
the free energy (\ref{eq:f0T}) in order to calculate the entropy
\begin{equation}
\mathcal{S}_{0}=-\left(\frac{\partial f_{0}}{\partial T}\right)_{h}=\ln\left(\frac{3+\sqrt{5}}{2}\right)\approx0.962423,\label{eq:S0}
\end{equation}
the sublattice magnetization of the Ising spins $M_{\mathrm{I},0}=-\left({\partial f_{0}}/{\partial h_{\mathrm{I}}}\right)_{T}={\sqrt{5}}/{10}$,
the sublattice magnetization of the Heisenberg spins $M_{\mathrm{H},0}=-\left({\partial f_{0}}/{\partial h_{\mathrm{H}}}\right)_{T}=1/2 \cdot\left(1+{\sqrt{5}}/{5}\right),$
while the total magnetization per unit cell equals
\begin{equation}
M_{\mathrm{t},0}=M_{\mathrm{I},0}+M_{\mathrm{H},0}=\frac{1}{2}+\frac{\sqrt{5}}{5}\approx0.9472136.\label{eq:mag-plt}
\end{equation}
This exact result will be confirmed later by numerical computation
at finite temperatures.

\begin{figure}[!t]
\centering \includegraphics[scale=0.95]{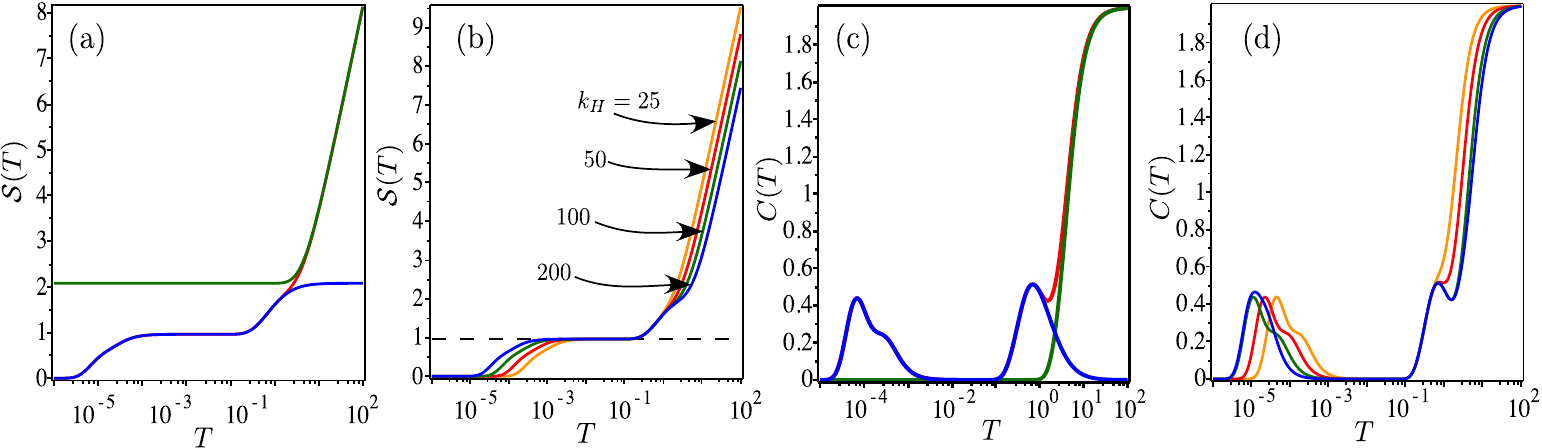}
\caption{(Colour online) \label{fig:Ent-Csp}~(a) The entropy as a function of temperature in
a semi-logarithmic scale for fixed values of $k_{\mathrm{H}}=100$, $\kappa=0.5$,
$J_{0}=-1$, $J_{x}=1$, $J_{z}=-1$, $\theta=\piup/2$ and $h=2$.
A blue line corresponds to the rigid model, a green line describes
the phonon contribution and a red line reports the overall entropy
of the model with the magnetoelastic coupling; (b) The overall entropy
as a function of temperature in a semi-logarithmic scale for fixed
values of $\kappa=0.5$, $J_{0}=-1$, $J_{x}=1$, $J_{z}=-1$, $\theta=\piup/2$,
$h=2$ and four different values of $k_{\mathrm{H}}=\{25,50,100,200\}$; (c)
Temperature variations of the specific heat corresponding to the panel
(a); (d) Temperature variations of the specific heat corresponding
to the panel (b).}
\end{figure}

Now, let us compare the magnetic behavior of the model accounting for
the magnetoelastic coupling in the vicinity of the triple point with that
of the fully rigid model in order to find out differences arising
from the magnetoelastic coupling. To this end, the entropy is plotted
in figure~\ref{fig:Ent-Csp}~(a) as a function of temperature in semi-logarithmic
scale, whereas a blue line corresponds to the fully rigid model, a
green line describes the phonon contribution and a red line reports
the overall entropy for the fixed values $k_{\mathrm{H}}=100$, $\kappa=0.5$,
$J_{0}=-1$, $J_{x}=1$, $J_{z}=-1$, $\theta=\piup/2$ and $h=2$.
It can be seen that the entropy closely follows the entropy of the
rigid model at low enough temperatures, whereas it tends to the phonon
contribution at a sufficiently high temperature. Figure~\ref{fig:Ent-Csp}~(c)
depicts the temperature dependencies of the specific heat corresponding
to figure~\ref{fig:Ent-Csp}~(a). Figure~\ref{fig:Ent-Csp}~(b) illustrates
the overall entropy of the model accounting for the magnetoelastic
coupling for the fixed values of $\kappa=0.5$, $J_{0}=-1$, $J_{x}=1$,
$J_{z}=-1$, $\theta=\piup/2$, $h=2$ and four different values of
the spring-stiffness constant $k_{\mathrm{H}}=\{25,50,100,200\}$. It is evident
from this figure that the entropy displays a plateau at $\mathcal{S}_{0}=\ln[{(3+\sqrt{5})}/{2}]$
regardless of the spring-stiffness constant in the range of moderate
temperatures $10^{-3}\lesssim T\lesssim10^{-1}$ before it tends to
zero in the asymptotic limit $T\to0$. Note that the fully rigid model
$(k_{\mathrm{H}}\rightarrow\infty)$ exhibits, for the considered set of parameters,
the residual entropy $\mathcal{S}_{0}=\ln[{(3+\sqrt{5})}/{2}]$,
which is however lifted for finite values of the spring-stiffness
constant due to the shift of the ground-state phase boundaries.
Finally, figure~\ref{fig:Ent-Csp}~(d) depicts temperature variations
of the specific heat corresponding to figure~\ref{fig:Ent-Csp}~(b),
where the formation of the additional small peak can be observed in
a low-temperature region due to the respective changes of the entropy
from a nonzero plateau to null.

\begin{figure}
\centering \includegraphics[scale=0.55]{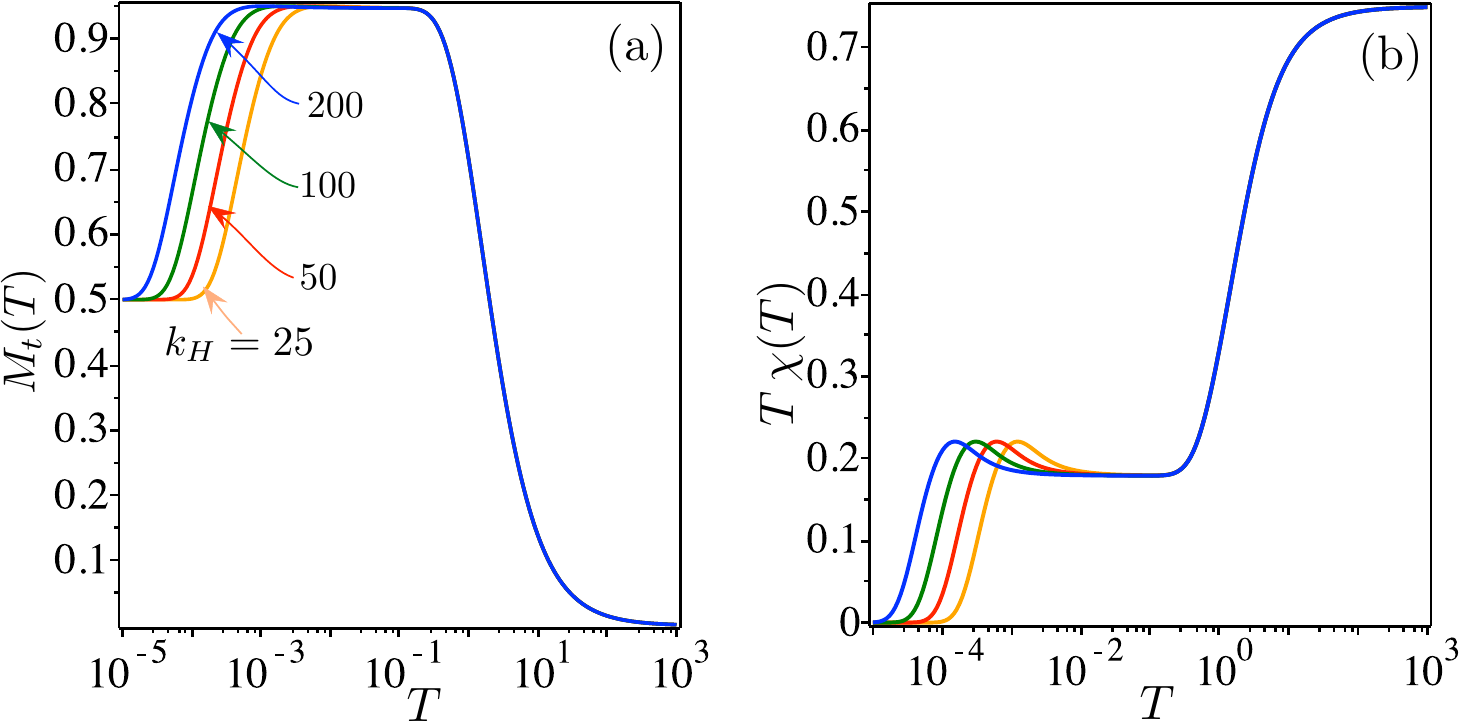}
\caption{(Colour online)\label{fig:Magn-Susc}~(a) The total magnetization per unit cell as
a function of temperature in a semi-logarithmic scale for fixed values
of $\kappa=0.5$, $J_{0}=-1$, $J_{x}=1$, $J_{z}=-1$, $\theta=\piup/2$,
$h=2$ and four different values of $k_{\mathrm{H}}=\{25,50,100,200\}$; (b)
A semi-logarithmic plot of the magnetic susceptibility times temperature
($\chi T$) product for the same parameter set as used in figure~\ref{fig:Magn-Susc}~(a)
for the magnetization.}
\end{figure}

The total magnetization is depicted in figure~\ref{fig:Magn-Susc}~(a)
against the temperature in a semi-logarithmic scale by assuming fixed
values of $\kappa=0.5$, $J_{0}=-1$, $J_{x}=1$, $J_{z}=-1$, $\theta=\piup/2$,
$h=2$ and four different values of the spring-stiffness constant
$k_{\mathrm{H}}=\{25,50,100,200\}$. It turns out that the total magnetization
per unit cell tends in the zero-temperature limit to the initial value
$M_{\mathrm{t}}=0.5$ irrespective of the spring-stiffness constant.  Then it
increases in agreement with the formula (\ref{eq:mag-plt}) to the
value $M_{\mathrm{t}}\sim0.947$ in the range of moderate temperatures $10^{-3}\lesssim T\lesssim10^{-1}$
before it finally tends to zero in the high-temperature region. Figure~\ref{fig:Magn-Susc}~(b)
illustrates the respective temperature variations of the magnetic
susceptibility times temperature ($\chi T$) product for the same
set of parameters as used in figure~\ref{fig:Magn-Susc}~(a) for the
magnetization. It is obvious that the product vanishes $\chi T\rightarrow0$
as temperature tends to zero, an intermediate plateau around the value
$\chi T\sim0.18$ is found at moderate temperatures $10^{-3}\lesssim T\lesssim10^{-1}$
and the product reaches the asymptotic value $\chi T\sim0.8$ in the
high-temperature limit.

\section{Conclusions}

In the present paper we have examined in detail the magnetoelastic
properties of the spin-1/2 Ising-Heisenberg diamond chain, which involves
two Heisenberg spins and one nodal Ising spin in each unit cell. It
is supposed that the decorating atoms involving the Heisenberg spins
harmonically vibrate perpendicular to the chain axis, while the nodal
atoms involving the Ising spins are placed at rigid lattice positions
when completely neglecting their lattice vibrations. In particular,
we have first investigated  the ground-state phase diagram depending
on the magnetoelastic constant and the spring-stiffness constant ascribed
to the Heisenberg coupling with the main emphasis laid on an investigation
of the parameter region close to a triple coexistence point of two
ferrimagnetic phases and one saturated paramagnetic phase. Next, we
have also examined in detail the thermodynamic properties at nonzero temperatures.

It has been found that the magnetoelastic nature of the Heisenberg
dimers is reflected through a nonzero plateau of the entropy in a
low-temperature region, whereas the specific heat displays an anomalous
peak slightly below the temperature region corresponding to the entropy
plateau. It also turns out that the magnetization exhibits a plateau
at almost saturated value in the same temperature region as the entropy
before it gradually tends to zero upon further increase of temperature.
The magnetic susceptibility displays within the plateau region an
inverse temperature dependence, which slightly drops above this plateau,
whereas an inverse temperature dependence is repeatedly recovered
at high enough temperatures.

\section*{Acknowledgments}

N. F. thanks Brazilian agency CAPES for full financial support.  J.
T. thanks Brazilian agency CNPq grant No. 159792/2019--3 for full partial
support. O. R. and S. M. de S thank CNPq and FAPEMIG for partial financial
support. J. S. thanks Slovak Research and Development Agency for financial
support provided under grant No. APVV--18--0197 and The Ministry of
Education, Science, Research and Sport of the Slovak Republic for
financial support provided under grant No. VEGA 1/0531/19. 

\ukrainianpart

\title{Магнетоеластичні властивості спін-1/2 ромбічного ланцюжка Ізінґа-Гайзенберґа поблизу потрійної точки співіснування}

\author{Н. Феррейра\refaddr{label1},
        Дж. Торріко\refaddr{label2}, С.М. де Соуза\refaddr{label1}, О. Рохас\refaddr{label1},  Й. Стречка\refaddr{label3}}

\addresses{
\addr{label1} Кафедра фізики, Федеральний університет Лавраса, C. P. 3037, 37200-900, Ліврас, Міна Жере, \\Бразилія
\addr{label2} Кафедра фізики, Федеральний університет Міна Жере, C. P. 702, 30123-970, Белу-Оризонті, \\Міна Жере, Бразилія
\addr{label3} Інститут фізики, Факультет природничих наук, Університет імені П. Й. Шафарика, парк Ангелінум 9, Кошиці 04001, Словаччина
}

\makeukrtitle

\begin{abstract}
Ми вивчаємо магнетоеластичні властивості спін-1/2 ромбічного ланцюжка Ізінґа-Гайзенберґа, елементарні комірки яких складаються з двох декорованих Гайзенборгових спінів і одного центрального Ізінґового спіна. Припускаємо, що кожна пара декорованих атомів, які несуть Гайзенборґові спіни, вібрує перпендикулярно до осі ланцюжка, в той час як Ізінґові спіни перебувають у фіксованих позиціях внаслідок відсутності інших вібрацій гратки. Вплив магнетоелестичної взаємодії на основний стан і скінченно-температурні властивості досліджуються біля потрійної точки співіснування в залежності від константи жорсткості, яка відноситься до Гайзенберґової взаємодії. Магнетоеластична природа Гайзенберґових димерів відображається через  ненульове плато ентропії в низькотемпературній області, в той час як теплоємність демонструє аномальний пік дещо нижче температурної області, що відповідає плато ентропії. Намагніченість також виявляє плато у цій же ж температурній області близько до значень насичення перед тим як вона поступово прямує до нуля зі зростанням температури. В межах області плато магнітна сприйнятливість показує обернену температурну залежність, яка руйнується нижче цього плато, в той час як обернена температурна залежність відновлюється при достатньо високих температурах.

\keywords магнетоеластичнй ланцюжок, спінова намагніченість, термодинаміка

\end{abstract}

\end{document}